\documentclass[tradiabstract]{aa}
\usepackage{graphicx}
\usepackage{txfonts}
\usepackage{natbib}
\def\HI{H{\,\small I}}
\newcommand{\msun}{{M$_\odot$}}
\newcommand{\mJybeam}{mJy beam$^{-1}$}
\newcommand{\mJyarcminsq}{mJy arcmin$^{-2}$}
\newcommand{\kms}{$\,$km$\,$s$^{-1}$}

\bibpunct{(}{)}{;}{a}{}{,}

\begin{document}

	\title{Recurrent radio emission and gas supply: the case of the radio galaxy B2~0258+35}

\titlerunning{Recurrent activity in B2~0258+35}
\authorrunning{Shulevski et al.}	
\author{Aleksandar Shulevski
	\inst{1,2}
	\and
	Raffaella Morganti
	\inst{1,2}
	\and
	Tom Oosterloo
	\inst{1,2}
	\and
	Christian Struve
	\inst{1,2,3}
	}
	\institute{Kapteyn Astronomical Institute, Rijksuniversiteit Groningen, Landleven 12, 9747 AD Groningen, The Netherlands\\
		   \email{shulevski@astro.rug.nl}
		   \and
		   Netherlands Institute for Radio Astronomy, Postbus 2, 7990 AA, Dwingeloo, The Netherlands\\
		   \email{morganti@astron.nl, oosterloo@astron.nl}
		   \and
		   Onsala Space Observatory, 439 92 Onsala, Sweden\\
		   \email{christian.struve@chalmers.se}
		  }
\date{\today}
	\abstract
	{Outlined is the discovery of a very faint, diffuse, low surface-brightness (0.5 \mJybeam, 1.4 \mJyarcminsq on average) structure around the radio source B2~0258+35 hosted by an HI-rich early-type galaxy (NGC~1167). Since B2~0258+35 is a young Compact Steep Spectrum (CSS) source, the newly discovered structure could represent a remnant from an earlier stage of AGN activity.
	We go on by explaining in detail all the possibilities for triggering the radio activity in B2 0258+35 regarding gas accretion in a recurrent AGN activity framework.
	NGC~1167 hosts a very regular, extended and massive \HI\ disc that has been studied in great detail. It has regular kinematics on large scales, which, together with stellar population studies of NGC~1167, exclude the possibility of a recent merger as the trigger for the current AGN activity responsible for the CSS source. Previous studies of the \HI\ closer to the core seem to go against the assumption of a circum-nuclear disc of \HI\ as the source of the accreting gas.
	We consider the cooling of gas from the hot, X-ray halo as a possible alternative option for the fueling of the AGN, as suggested in the case of other sources of similar radio power as B2~0258+35. This would provide a more likely explanation for the recurrent activity. Further, if the suggestion made by Giroletti et al.(2005) that the inner CSS may not be able to grow to large scales is correct, this implies that different cycles of activity may have different characteristics (e.g. radio power of the emission).
	Estimates are given for the age of the faint diffuse emission as well as for the current accretion rate, which are in good agreement with literature values.
	If our assumptions about the accretion mechanism are correct, similar large-scale, relic-like structures should be more commonly found around early-type galaxies and this will be hopefully confirmed by the next generation of sensitive, low-frequency radio surveys.
	}

\keywords{galaxies: active - radio continuum: galaxies - galaxies: individual: B2 0258+35}	
\maketitle

\section{Introduction}
\label{sec:intro}
	
Active Galactic Nuclei (AGN) have been recognized in recent years to have a profound influence on their surrounding interstellar medium (ISM) and, as consequence, also  on the evolution of the host galaxy \citep[see][]{RefWorks:91, RefWorks:92, RefWorks:55}.
Radio-loud AGN can exert this influence not only through their collimated radio jets but also through the cocoon of shocked medium around them \citep[][]{RefWorks:56, wag}. Therefore, they can influence a large volume in (and outside) the host galaxy. 
Although this type of AGN is relatively rare and shortlived, the radio phase can be recurrent during the life of  the galaxy \citep[see][for some examples]{RefWorks:6, RefWorks:45, RefWorks:42}. This may increase substantially the impact that radio-loud AGN have on the ISM and their role in galaxy evolution. However, the  life-cycle of a radio source has still many open questions (e.g. is radio activity occurring in every galaxy and what is the "duty cycle" of the recurrent activity) limiting our understanding of the impact of this type of nuclear activity. 

What do we know about the life-cycle of a radio source? The first phase of a radio source has been identified with compact sources with steep or peaked spectrum (so called Compact Steep Spectrum (CSS) and GigaHertz Peaked Spectrum (GPS) sources), see \cite{RefWorks:57}, \cite{RefWorks:58} and \cite{RefWorks:31} for a recent overview. These sources have already the morphology of grown-up sources but their size is comparable to galactic scales, i.e. the inner few kpc of the host galaxy.  Most of them are expected to grow to large radio galaxies \citep{RefWorks:31} although some  may actually never reach this phase, either because the fueling of the AGN "engine" stops or because of a hostile ISM \citep[][]{RefWorks:93}. See also \cite{RefWorks:104} and \cite{RefWorks:94}, Chapter 6.

The typical lifetime of a radio source ranges between $ 10^7 $ and $ 10^8 $ yrs \citep{RefWorks:83, RefWorks:96} and after that the nucleus switches off. This may result in the formation of a relic source that will slowly fade away. The relic structures (with no nuclear activity present) are very rare, with only an handful known \citep{RefWorks:97, RefWorks:96}. More common seems to be the situation where the  radio source is active intermittently; in that case one may find fossil radio plasma left over from an earlier phase of activity, while newly restarted core and radio jets are visible as well.
Evidence for a re-start in the activity of radio sources, after a period of shut down of the central engine, or of rejuvenated sources has been found in a number of cases \citep[][]{RefWorks:40, RefWorks:34, RefWorks:3, RefWorks:6} although proper statistics of the occurrence and characteristics are not available.

The off phase appears to be in most of the cases shorter than or at most comparable with the active phase \citep{RefWorks:96}. However, statistical studies using luminosity functions \citep{RefWorks:43, RefWorks:42} have shown that the life-cycle of radio loud AGN depends on the radio power, with powerful (i.e. Fanaroff-Riley type 2 or  FRII, \citealp*{RefWorks:12}) radio sources becoming active only every one-to-few Gyr while low power radio sources (Fanaroff-Riley type 1, FRI) would need to spend over a quarter of their life in an active phase \citep{RefWorks:42}. This would suggest that signs of past radio-loud activity could be more common in the latter sources. Unfortunately, no systematic search for such signatures has been possible so far and studies of single objects are relatively sparse and limited \citep[][]{RefWorks:3, RefWorks:51}. Whether radio emission from a previous phase of activity is still observable, it also depends on the external conditions, with the hot, X-ray environment particularly suitable for keeping the relic confined and limiting the adiabatic losses \citep[][]{RefWorks:34}.

For understanding the origin of the recurrent nuclear activity, it is also important to have comprehensive information about the assembly and evolution of the host galaxy. The commonly considered way to (re-)trigger an AGN is to provide fresh supply of gas. The presence of gas is often observed in early-type galaxies, typical hosts of radio-loud AGN. Although the presence of this gas does not appear to have a clear connection with the presence of an active nucleus \citep[see][]{RefWorks:67}, it is interesting to note that the occurence of \HI\ in restarted sources seems to be higher than in other radio sources \citep[][]{RefWorks:89, RefWorks:6}. This could suggest a possible link between the presence or injection of gas and the activity.

Thus, identifying cases of recurrent radio activity and understanding their time scale (as well as the gas content and kinematics close to the AGN) is challenging but extremely important in order to fully probe the impact of radio-loud AGN on the host galaxy and their importance for feedback effects. The study of objects for which information at different wavebands is available allows a better understanding of the origin of the activity and to connect the history of the host galaxy with the history of the nuclear activity. This combination has triggered our interest in the object which is the subject of this paper.\\

In this paper we present the discovery of diffuse, low brightness extended radio emission around the young CSS radio source: B2~0258+35. A hint of the presence of this structure was first detected in a preliminary continuum image obtained by \cite{RefWorks:13}. This has inspired a more detailed look at the data that we present now in this paper.\\

B2~0258+35 is hosted by the field, early-type galaxy  NGC~1167 ($z = 0.0165$\footnote{The adopted cosmology in this work is: $  H_{0} = {73} $ \kms Mpc$^{-1}$ , $ \Omega_{matter} =  0.27 $, $ \Omega_{vacuum}  =  0.73$. At the redshift of B2~0258+35, $1^{\prime\prime} =  0.34$  kpc}). 
The central radio source has been studied by \cite{RefWorks:15} and classified as a CSS source. These authors derived an age for this structure of $ 9 \cdot 10 ^{5}$ yr. The CSS source has a radio luminosity of $ L _{408 \, {\rm MHz}}  = 10 ^{24.37}$ WHz$^{-1} $. The radio structure of the CSS source does not have hot spots (despite the fact that the structure appears to be the result of a strong interaction with the ISM), so if this is indeed a young radio galaxy, it might evolve into a FR I. This is consistent with the measured radio power. However, \cite{RefWorks:15} have argued that this source might represent an example of a CSS source that will not grow to become a kilo-parsec scale radio galaxy.  

What makes B2~0258+35 notable is the presence of {\sl a large, massive disk of \HI\ } that has been studied in detail and can give us an additional insight into the formation history of the host galaxy. The disk (with M$_{\HI}= 1.5 \cdot 10 ^{10}$ \msun\  and diameter of 160~kpc, see Fig.1 right, \citealp*{RefWorks:49} and \citealp*{RefWorks:25}), shows {\sl extremely regular kinematics} within the inner $ r < 65$ kpc and signs of interaction with several satellite galaxies in the outer regions where the gas appears slightly disturbed. The detailed work of \cite{RefWorks:25} shows that the disc has grown by accretion of cold gas from satellite galaxies. Furthermore, its regularity implies that the host galaxy has not suffered a major merger in the center in at least the past  1~Gyr. \HI\ has been detected also in absorption - with much higher resolution observations - against the central CSS source. The kinematics of this gas have been studied in \cite{RefWorks:13}, Chapter 7, and appear quite regular, consistent with the velocities of the large-scale disc. However, a blue-shifted, possibly outflowing, component has being detected both in \HI\ and in CO. Thus, thanks to the \HI\ we have a clear view of the recent assembly history of the host galaxy. In this paper we explore how this relates (or not) to the radio-loud phase(s) of activity.\\
	
The paper is structured in the following way. In Section \ref{sec:data} we describe the observations and the data reduction procedure. Section \ref{sec:relic} presents the results and discusses the origin of the newly found extended radio structure as a possible signature of recurrent activity in this galaxy. In Section \ref{sec:gas} we combine this with information about the gas supply by using results from (recent) \HI\ studies. Section \ref{sec:fin} presents some additional implications and discusses further work prospects.

\begin{figure*}[htpl]
\centering
\begin{minipage}[c]{0.5\linewidth}
\centering \includegraphics[width=100mm]{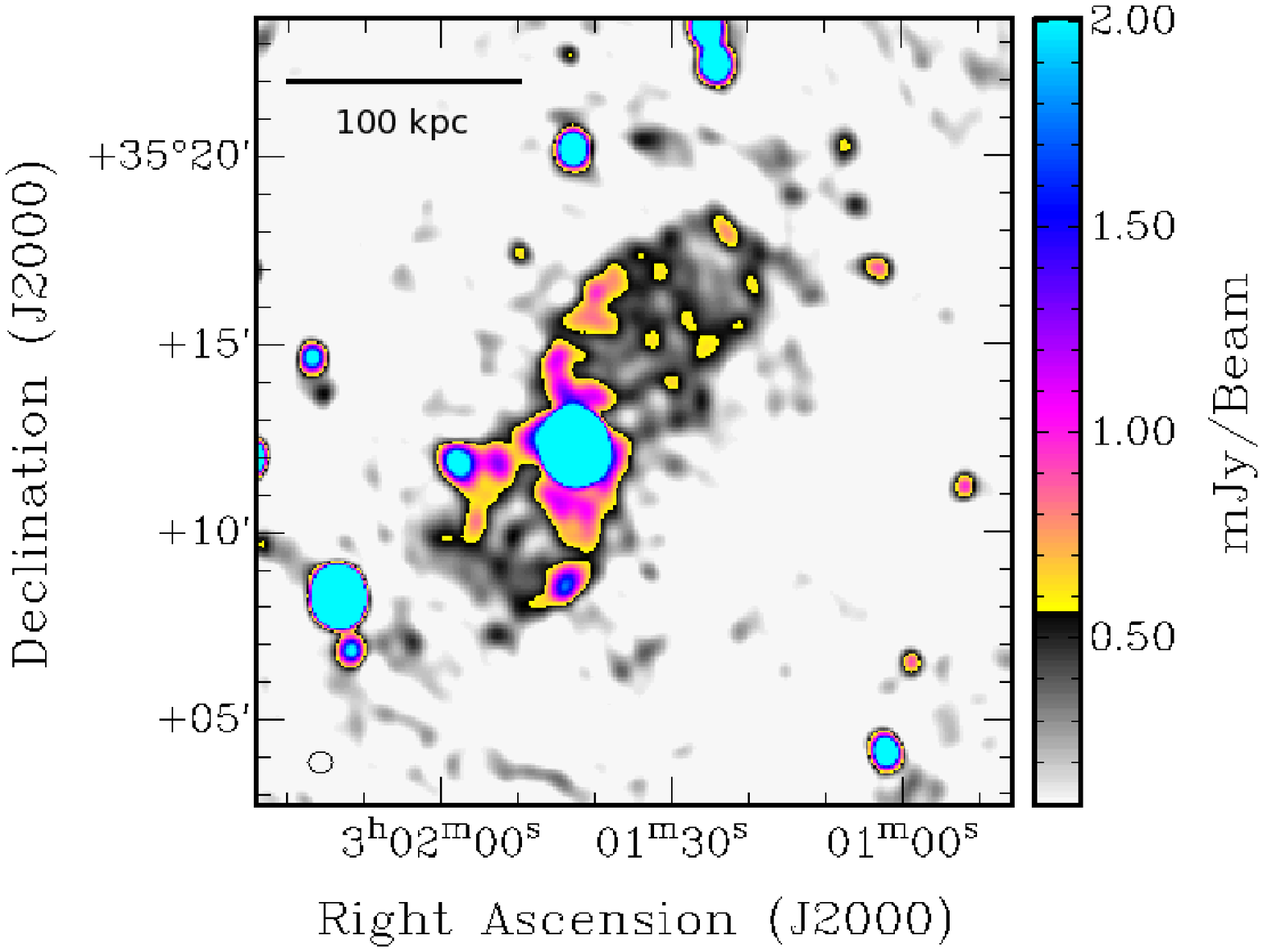}
\end{minipage}%
\begin{minipage}[c]{0.5\linewidth}
\centering \includegraphics[width=80mm]{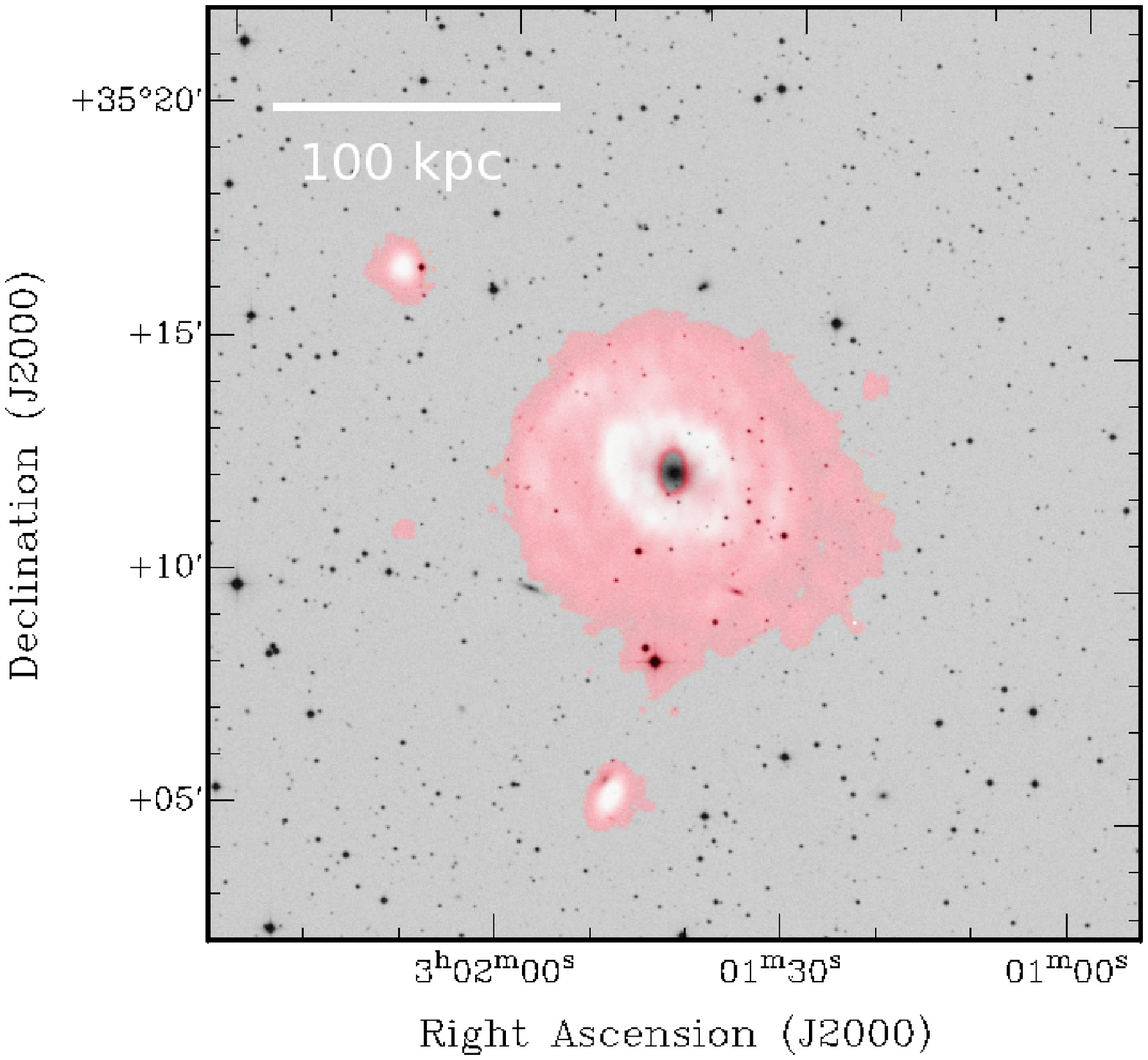}
\end{minipage}
\caption{Left: continuum image  of the diffuse emission around B2 0258+35. The synthesized  beam  is indicated with the ellipse at bottom left. The intensity ranges from 100 $ \mu Jy $ ($ 1 \sigma $) to 2 mJy ($ 20 \sigma $).  Right:  21-cm \HI\ total intensity image taken from \cite{RefWorks:13} superposed onto a DSS2 optical image. Scale as indicated.}
\label{map}
\end{figure*}

\section{Observations and Data Reduction}
\label{sec:data}

B2~0258+35 was observed with the Westerbork Synthesis Radio Telescope (WSRT) for 12 x 12h. The data were originally taken for line (\HI) observations (analysis and discussion of the results are presented in \citealp*{RefWorks:25}; see that paper for further details). Because of this, the observations were centered around the frequency of 1.39 GHz (\HI\ in the rest frame of the source) and used a bandwidth of 20 MHz. The data reduction was done using the {\sc Miriad} package \citep{RefWorks:63} as described in \cite{RefWorks:25}. 3C~48 and 3C~147 were used as flux and band-pass calibrators respectively.

We have extracted from these data the continuum, using the line-free channels in the band. The continuum subtraction was done on the visibilities using the task {\sl uvlin} with a second order polynomial. Data from eleven of these observing runs were used to produce continuum maps. One data-set was not included in the final map because of low quality of the data due to technical problems in that observation. 

Each observation was separately imaged and several self-calibration runs were performed (solving, at the beginning, for phase only and for amplitude and phase at the end of the procedure) before obtaining the final image. The images were obtained using uniform weighting, but were subsequently convolved with a gaussian of 30$^{\prime\prime}$ FWHM to enhance the extended emission. These images were then combined together to produce a final image where an extended low surface brightness emission appears very clear (see Fig. \ref{map}, left panel). The r.m.s. noise of the final image is 100 $\mu$Jy  beam$^{-1}$, with a beam size of $ 39^{\prime\prime} \times 33^{\prime\prime}$ at a position angle of $ 2^{\circ} $.

\section{Low surface-brightness, extended structure: a radio relic?}
\label{sec:relic}

The final radio continuum image is shown in Figure \ref{map} in the left panel. In addition to the central (unresolved at our spatial resolution) CSS source, we detect an extended, low surface brightness structure. The average surface brightness is 0.5 \mJybeam (measured in a region encompassing the northern lobe). This translates to around 1.4 \mJyarcminsq which is on the faint end compared to the source sample published by \cite{RefWorks:101}. Thus, even after the spatial smoothing used, the structure remains very faint, only a few sigma above the noise. The large contrast in the flux levels between the central CSS source and the extended structure makes it extra complicated to image this structure. The extended radio emission has a total projected linear size of 240 kpc and it has a distinctive double-lobed appearance, with, in some locations, relatively bright edges. The central CSS source has a total flux of 1.8 Jy ($ L _{1.4 \rm GHz}  =  2.1  \cdot \, 10 ^{23}$  WHz$ ^{-1} $), while the peak of the extended structure is $\sim 2.4$ mJy. It has a total flux of 119 mJy (integrated over a region determined by visual inspection of the image encompassing the entire diffuse emission region). This gives a radio luminosity of  $ L _{1.4 \rm GHz}  =  5.5  \cdot \, 10 ^{22}$  WHz$ ^{-1} $ for the  diffuse emission. 

\begin{figure}[htpl]
        \centering
        \includegraphics[width=90mm]{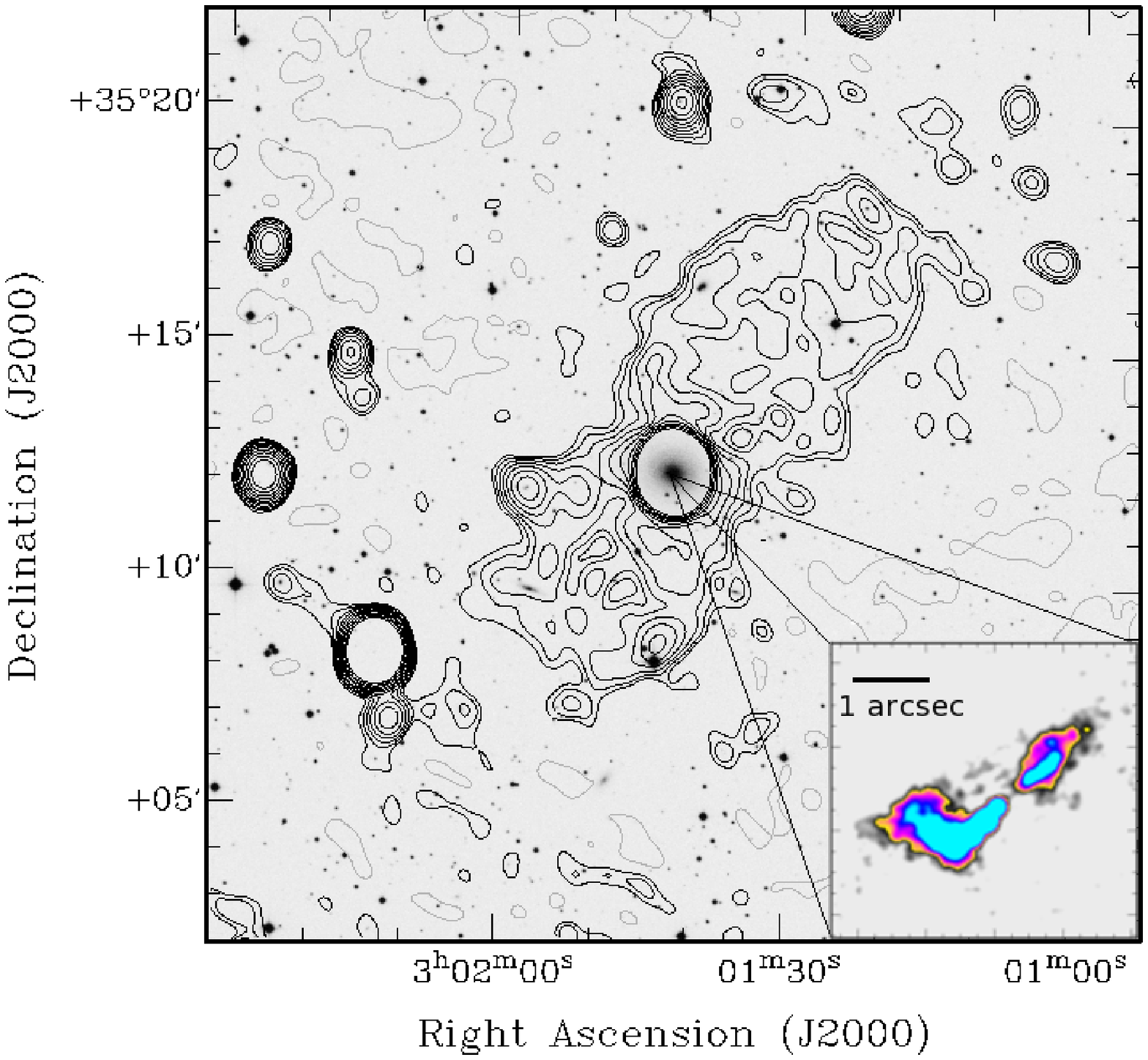}
	\caption{Contours of the diffuse radio emission around B2 0258+35 overlaid on a DSS2 image. The contour levels range from 0.2 \mJybeam to 20 \mJybeam ($ 2 \sigma $ to $ 20 \sigma $) increasing by a factor of 1.5 and are marked in black. Negative countours at -0.2 \mJybeam are gray. The inset at bottom right shows the VLA image of the central CSS source \citep{RefWorks:15}. Note the similarity in the orientation between the CSS source and the diffuse emission.}
	\label{over}
\end{figure}

Because of the extremely low surface brightness and the lack of collimated and/or compact features (like jets and hot spots), the newly found, extended radio structure can be explained as  lobes left over from a previous cycle of nuclear activity. The spatial resolution of our image does not allow to verify whether a connection exists between the inner CSS source and the extended, low-surface brightness lobes. We also cannot derive any value for the spectral index: if some injection of "fresh" electrons is still on-going, we would not expect these extended structures to have an extremely steep spectral index. Thus, the issue of whether the extended, low-surface brightness structure in B2~0258+35 is a relic structure or the lobes still get some feeble injection of fresh electrons cannot be answered with the present observations. This will be done with already planned follow-up, low frequency observations.\\

However, it is worth noting that the extended structure shows relatively bright edges, in particular in the northern lobe. This could be the result of the interaction of the radio plasma with the ISM/IGM and may suggest in-situ acceleration of the electrons. Alternatively, the structure is not a relic but radio plasma is still flowing through from the active nucleus, albeit likely at low rate. The morphology of the extended structure bears some similarities with radio-bubble structures found in low-luminosity radio sources (see \citealp*{RefWorks:64} for a compilation) and in some radio galaxies like Centaurus~A \citep{RefWorks:60} and M~87 \citep{RefWorks:65}. In both cases, spectral index studies (\citealp*{RefWorks:61} for Centaurus~A and the recent study at low frequencies using LOFAR for M~87, de Gasperin et al. 2012, to be submitted) have found no steepening in the spectral index of the diffuse lobes. This suggests that, indeed, some injection of fresh electrons is still ongoing.\\

In the absence of spectral index data,  we attempt to obtain an estimate of the age of the extended radio emission in B2~0258+35 by treating it as buoyant bubbles rising through the IGM away from the host galaxy. The forces considered acting on the bubble are the buoyancy and IGM ram pressure forces. Under these assumptions, we can use an expression for the velocity with which the bubble rises \citep{RefWorks:86, RefWorks:65}:

\[ v \sim \sqrt{\frac{r}{R} \frac{8}{3C}} \, \cdot \, v_{k} \]

\noindent where $ r $ is the radius of the bubble, $ R $ the distance of the bubble center from the host galaxy, $ v_{k} \, = \, \sqrt{gR} $ is the Keplerian velocity at the distance of the bubble, $g$ the gravitational acceleration and $ C $ is the drag coefficient. We restrict our attention to the northern lobe and we approximate its radius (assuming spherical geometry) by half of its width (as observed in Figure \ref{map}) for our purposes. This is motivated by the fact that the width would reflect most closely its real size; the height is influenced by the expanding motion, so it would give us a worse size estimate. So, we have: $ r \, = \, 50$ kpc, $ R  =  50$  kpc and $ C \, = \, 0.75 $. The value for the drag coefficient is adopted by following \cite{RefWorks:65} where they arrive at it based on hydro-dynamical simulations of buoyant bubbles traveling through a stratified and compressible medium. In contrast, \cite{RefWorks:86} adopt a value of $ 0.5 $, which is a value for a solid sphere moving through an in-compressible fluid. We can take a value for the Keplerian velocity from the work of \cite{RefWorks:25} who measured the rotational velocity of the \HI\ disc at a distance of $ R \, = \, 50 \, kpc $ from the host galaxy to be $ v_{k} \, = \, 325 $ \kms. Using the above mentioned values for the variables, we obtain for the speed with which the bubble rises a value of $\sim  613 $ \kms. Treating this bubble as originating from the last burst of AGN emission before shutdown, and assuming a uniform motion, we derive the time elapsed since the activity ceased to be $\sim 8 \cdot 10^{7} $ yr. This is a lower limit because of projection effects, and it is in broad agreement with estimates from publications cited in this paper for relics around CSS sources.\\

It is interesting to note that the estimated age of the CSS source is very small (the ratio of the timescales is on the order of $ 0.01 $) when compared to the estimated time elapsed since cessation of the AGN activity in the previous cycle. Thus, if we assume that the CSS source marks the beginning of the current, newest, phase of AGN activity, the time between subsequent phases of activity is on the order of $ 10^{8} $ yrs, in agreement with current estimates \citep{RefWorks:96}. All of this holds, of course, if we assume that the observed extended structure is indeed a radio relic.

\section{Recurrent activity and supply of gas}
\label{sec:gas}

As was mentioned in the Introduction, radio structures that represent signatures of past activity in the host galaxy have been found in a number of objects. The question for all these objects - including  B2~0258+35 - is what causes the activity to stop and restart. Gas and dust are considered the source of fuel for the triggering (or re-triggering) of AGN. Although radio-loud AGN are typically hosted by early-type galaxies, it has been shown that  gas is observed in a large number of them at least on the kpc scale \citep[][]{RefWorks:67, RefWorks:87, RefWorks:68, RefWorks:88}. Therefore, the availability of gas, at least on these large scales, is not a problem. However, the presence of this gas  does not appear to have a clear connection with the presence of an active nucleus \citep{RefWorks:67}. 

Perturbations produced by mergers or accretions could provide a mechanism for the gas to loose angular momentum and fall into the Super Massive Black Hole (SMBH). B2~0258+35 is a rare case where the recent merger history of the host galaxy can be reconstructed from the kinematics of the gas. \HI\ has been detected in this object not only in absorption against the radio core, but also in a large-scale ($\sim 160$ kpc diameter) disk structure (see Fig. 1, right panel). Due to sensitivity limitations of current radio telescopes, \HI\ emission is quite rarely detected at the typical distance of radio galaxies.  Even more rare are cases of large \HI\ disks where a detailed analysis of the kinematics of the gas can be done. The kinematics of the \HI\ over the inner 65 kpc radius are extremely regular (see Fig. 5 in \citealp*{RefWorks:25}). Outside this radius, signs of recent or ongoing interaction are seen, although the \HI\ disk remains kinematically regular. This has suggested that, even if part of the gas was brought in by a major merger, this must have been {\sl more than a Gyr ago}. After that, the continuous supply from gas-rich satellite galaxies has been the main mechanism bringing new cold gas. However, this has happened at the outskirts of the disk and without significantly disturbing the kinematics of the gas. Small halo clouds were not found in this object.\\

Thus, considering this detailed analysis and comparing the time scale with those of the radio structures derived for the "relic" (Section \ref{sec:relic}) and for the inner CSS \citep{RefWorks:15}, we conclude that  despite the presence of a huge reservoir of \HI\ in this object, we do not find a link between merger or accretion activity and the cycle of radio emission. The data rule out a major merger as trigger of the radio emission. Signatures of minor accretions are observed in the outer regions of the disk and they do not appear to have left any signature that could indicate a clear link between  the formation history of the host galaxy and the life-cycle of the radio activity. Even invoking an extreme time-delay between these events (larger than in other studied cases, see \citealp*{RefWorks:74}), the fact that more than one stage of activity is seen cannot be unambiguously linked to  clear events in the process of building of the gas disk. This is reminiscent of what was already found in the case of Centaurus A \citep{RefWorks:98}. Therefore, despite the large reservoir of \HI\ in the disk, the actual fuel of the AGN may have a different origin.\\

Mass loss from stars formed during a star-burst phase (e.g. triggered by a merger) has been suggested as a reservoir for the growth of the black hole and the trigger of its activity \citep{RefWorks:71}. This way of fueling could also explain a delay (of a few $ \times $ 100 Myr) between the star-burst phase and the onset of the radio phase observed in a number of radio sources \citep[][]{RefWorks:74}. However, in the case of B2~0258+35, the optical spectrum does not show signatures of young stars \citep[][Chapter 4]{RefWorks:73}. This would imply a far too large delay (of the order of Gyr) between the stellar evolution and the onset of the radio emission, making this scenario very unlikely.\\

The possibility of having warm, halo clouds ($ 10^{4} \, - \, 10^{6} $ \msun) fueling the AGN has been recently proposed by \cite{RefWorks:44}. The main problem in this scenario is the assumption that the orbits of these warm halo clouds are random. However, there is clear evidence that the halo clouds are likely rotating, and are not in random orbits. Any amount of rotation would cause these clouds never to fall to the center (only the non-rotating clouds on radial orbits will fall to the center) but, instead, to strongly interact with the halo (\citealp*{RefWorks:99}). The current theory is also that most of these clouds would evaporate before hitting the disk or the center \citep{RefWorks:100}.\\

Finally, the possibility that gas cooling from the hot galactic halo supplies fuel to the SMBH via e.g. Bondi accretion, and is a dominant mode of accretion in low-power radio galaxies  has been suggested by a number of studies \citep[][]{RefWorks:70, RefWorks:75, RefWorks:62, RefWorks:95}. This is also supported by the analysis of the optical emission lines in radio galaxies carried out by \cite{RefWorks:90}. In radio galaxies with low-excitation spectra (a group that includes all the low-power radio galaxies, i.e. FR I type) the characteristics of the optical lines and the powering of the jets can be explained as proceeding directly from the hot, X-ray emitting phase of the ISM/IGM in a manner analogous to Bondi accretion. In addition to this, \cite{RefWorks:102} have suggested a possible formation mechanism for high-velocity clouds; they can in turn supply the fuel to the SMBH.\\

Using the information from the radio data, we can infer the  accretion rate necessary for fueling the radio AGN following the method presented by \cite{RefWorks:75}, \cite{RefWorks:95} \citep[see also][]{RefWorks:77}. The radio core flux density of B2 0258+35 at 1.6 GHz was measured by \cite{RefWorks:15} - using the VLBA - to be $ 7.4 \, mJy $.

This gives a core luminosity of $ L_{core} \, \approx \, 3.5 \, \cdot \, 10^{21} \, W \cdot Hz^{-1} $, which can be converted to jet power - $ P_{j} \, \approx \, 1.2 \, \cdot \, 10^{36} \, W $. \cite{RefWorks:95} found a relation between the Bondi accretion power and the jet power that, if applied to B2~0258+35 gives $ P_{B} \, \approx \, 6.5 \, \cdot \, 10^{37} \, W $, corresponding (using: $ P_{B} \, = \, 0.1 \dot{M} c^{2} $ ) to a  mass accretion rate of at most $ 0.1 $ \msun\ $ yr^{-1} $.

Unfortunately, although X-ray observations at 2-10 keV \citep{RefWorks:76} show a X-ray halo which extends beyond $D_{25} \, \sim \, 49 \, kpc$ (Rasmussen, private communication), the modeling of the density and temperature profile for estimating the Bondi accretion rate is not available. Thus, we limit ourselves to comparing the values obtained with what is found for other radio galaxies. For example, using the same reasoning as outlined above, $\dot M \sim 0.1$  \msun\ $ yr^{-1} $ was found for NGC~315 \citep{RefWorks:77}, quite high when compared to e.g. the sample in \cite{RefWorks:75} but inside the range (although toward the high end) of the distribution for radio galaxies found by \cite{RefWorks:95}. This seems to support the argument that cooling of hot gas, resulting in a Bondi - type of accretion can be responsible for feeding the AGN.\\
 
If the cooling of the hot gas is at the origin of the fueling of the activity, one may wonder whether, in the process of cooling, the gas would spend enough time in the \HI\ phase, and be observable in \HI\ absorption against the nuclear regions of radio continuum. The advantage of observations of \HI\ in  absorption is that the gas can be traced to very small scales (unlike for hot gas observed in X-ray) and can provide the kinematics and the distribution of at least the cold component. In this respect, it is intriguing that the occurrence of \HI\ in the central regions of restarted radio sources appears to be higher than in other radio sources (\cite{RefWorks:89}, \cite{RefWorks:6}, Gereb et al. in prep.). If confirmed for more objects, this suggests that the presence/availability of (cold) gas may be connected with the duty-cycle of radio sources. B2~0258+35 confirms this trend with \HI\ absorption detected against the central component \citep{RefWorks:13}. However, is this \HI\ able to reach the SMBH and fuel the AGN?\\
 
In the case of the FR I radio galaxy NGC~315, an in-falling cloud of \HI\ has been observed at a few pc distance from the core \citep{RefWorks:77}, corresponding to accretion rates in the range $10^{-4} - 10^{-3}$ M$_{\odot}$ yr$^{-1}$ (inferred from \HI\ absorption studies using the radio continuum of the core). Although the presence of the \HI\ in-fall is intriguing, the accretion rate  is lower than what is required to fuel the AGN. However, it has to be mentioned that this is probably a lower limit, since the \HI\ absorption probes only the gas detected against the continuum. In the case of B2~0258+35, the highest resolution available so far for \HI\ absorption studies does not reach such small linear scales. The \HI\  absorption study \citep{RefWorks:13} shows that, down to a linear scale of $\sim 300$ pc, most of the gas has relatively regular kinematics, in good agreement with the kinematics of the large scale disc. At this resolution, the only deviation from circular motion appears to be the gas associated with a blue-shifted, possibly out-flowing, component that has been detected both in \HI\ and in CO.  If the lack of further deviation from circular motion, and in particular in-falling cloud(s), is confirmed by high resolution VLBI observations, this would imply that the hot gas dominates the accretion and e.g. the cooling gas would spend only a very short time in the \HI\ phase.

If the activity is related to cooling of the X-ray gas, one can imagine, as already described by other authors, a cycle that would explain not only the triggering but also the interruption of the activity.

The radio emission would be responsible for heating up the ISM/IGM, thus stopping the accretion, and, as a consequence of this, the radio emission would be interrupted. This would allow the ISM/IGM to cool again, and after a while the radio emission would restart - thus the cycle can go on \citep[][]{RefWorks:78, RefWorks:45, RefWorks:103}.
 
\section{Conclusion and Future Studies}
\label{sec:fin}

We have presented the discovery of an extended and low-surface brightness radio emission around the young (CSS) radio source B2~0258+35 hosted in an \HI-rich early-type galaxy. The newly found radio emission has a distinct double-lobed appearance and is likely a left over from a previous cycle of activity of the galaxy. The faint lobes may actually not be completely dead relics but still weakly refueled with "fresh" electrons from the nucleus in a similar way as e.g. the middle lobe of Centaurus~A \citep{RefWorks:60}. This may explain why the large and faint structure is still visible (despite the source not being in a cluster environment). Only a study of the spectral index may help in verifying this hypothesis.

It is intriguing to note the speculation from \cite{RefWorks:15} that the inner CSS (with an estimated age of $ 9 \, \cdot \, 10^{5} $ yr) might not grow to become a large-scale radio galaxy. No final hot spots are observed although the structure appears to be the result of a strong interaction with the ISM. Considering that the previous cycle of activity did manage to expand to hundreds of kpc in size, this would suggest that every cycle of activity can have different characteristics or that the ISM is now particularly rich to affect more drastically the new radio source.  

The huge \HI\ disk found in B2~0258+35 has a very regular kinematics that do not show  any obvious connection with the triggering of the radio source. Following the conclusion of \cite{RefWorks:25}, no major merger has recently occurred in this galaxy and, therefore, rules out a major merger as a trigger of the radio emission. Minor accretions are also ruled out, because they appear only to have mildly perturbed the \HI\ at large radii. We do not see indication in the high resolution data \citep{RefWorks:13} that (some of the) gas in a circum-nuclear disc is actually fueling the AGN. However, we cannot completely exclude the presence of such a gas until even higher resolution data are available. Future VLBI study may be able to shed more light on this.
The possibility that, despite the large reservoir of \HI\ in the disk, the actual fuel of the AGN may originate from the cooling of gas in the hot halo is an interesting alternative and would also help to explain the recurrent activity.

In the future, sensitive low-frequency surveys that cover large area of the sky will allow finding out how common is the presence of radio emission left over from previous cycle of activity and tell us about the duty cycle of the radio emission. Sensitive, low-frequency observations are a robust way to identify these sources, see \cite{RefWorks:2}, \cite{RefWorks:54}. Increasing  the known number of such structures (which is now limited) will allow us (if combined with multi-waveband information) to understand under which conditions the radio phase (re-)starts, the time-scales of this phenomenon and, as a result, learn more about the impact of the radio plasma.
In particular, if the cooling of the hot halo is, indeed, a reservoir for the triggering of the AGN, then we would expect relatively regular cycles of activity and we would expect to observe structures similar to B2~0258+35 in many more objects.

All these requirements are currently being met by LOFAR (and will be met by the SKA in the future). The deep surveys planned with LOFAR \citep{rott10} that will cover large areas of the sky will allow us to find how common are structures like the one detected in B2~0258+35. LOFAR 60~MHz observations of B2~0258+35 have been already performed in order to detect or set a limit to the brightness of the extended structure at lower frequency and learn more about its origin. These data  will be presented in a future paper.

\begin{acknowledgements}

Data taken with the Westerbork Synthesis Radio Telescope were the basis of this work. It is operated by ASTRON (Netherlands Institute for Radio Astronomy) with support from the Netherlands Foundation for Scientific Research (NWO).\\
This research has made use of the NASA/IPAC Extragalactic Database (NED) which is operated by the Jet Propulsion Laboratory, California Institute of Technology, under contract with the National Aeronautics and Space Administration. 

\end{acknowledgements}

{}

\end{document}